\documentclass[preprint]{aastex}

\def\simle{\,\hbox{\hbox{$ < $}\kern -0.8em \lower 1.0ex\hbox{$\sim$}}\,}

\def\msun{$M_{\odot}$}
\shortauthors{Thorstenen et al.}
\shorttitle{Cataclysmic Binary V405 Peg (RBS 1955)}
\begin{document}
\title{V405 Peg (RBS 1955): A Nearby, Low-Luminosity Cataclysmic Binary
\footnote{Based in part on
observations obtained at the MDM Observatory, operated by
Dartmouth College, Columbia University, Ohio State University, and
the University of Michigan.}
}

\author{John R. Thorstensen}
\affil{Department of Physics and Astronomy,
6127 Wilder Laboratory, Dartmouth College,
Hanover, NH 03755-3528, USA;
john.thorstensen@dartmouth.edu}

\author{Robert Schwarz, Axel D. Schwope, A. Staude, J. Vogel, M. Krumpe, J. Kohnert, A. Nebot G\'omez-Mor\'an}

\affil{Astrophysikalisches Institut Potsdam (AIP), An der 
Sternwarte 16, 14482 Potsdam, Germany}

\begin{abstract}

The cataclysmic binary V405 Peg, originally discovered as ROSAT 
Bright Source (RBS) 1955 (= 1RXS J230949.6+213523), 
shows a strong contribution from a late-type secondary star in its
optical spectrum.  This led \citet{rbsid}
to suggest that it is among the nearest cataclysmic binaries.
Here we present extensive optical observations of V405 Peg.
Time-series spectroscopy shows the orbital period, $P_{\rm orb}$,
to be $0.1776469(7)$ d (= 4.2635 hr), or 5.629 cycles d$^{-1}$.  
We classify the secondary as M3 - M4.5, and estimate its contribution
to the light.  Astrometry with the MDM 2.4m telescope gives a parallax
$\pi_{\rm abs} = 7.2 \pm 1.1$ milli-arcsec, and a relative proper motion of
$58 \pm 1$ mas yr$^{-1}$.
We employ a Bayesian technique to form a best estimate of the 
distance from the parallax and other indicators; this 
yields $d = 149 (+26, -20)$ pc.  
The semiamplitude of
the secondary stars's radial velocity is $K_2 = 92 \pm 3$ km s$^{-1}$,
indicating a fairly low orbital inclination if the masses are typical.
Extensive time-series observations in the $I$-band show the system 
varying between a minimum brightness level of $I=14\fm 14$ and 
states of enhanced activity about 0.2 mag brighter.
During the low state, the light curve is dominated by 
ellipsoidal modulation of the secondary star 
at a period consistent with the spectroscopic value. 
In the high-state we detect an additional photometric 
modulation with $\sim$ 0.1~mag amplitude, which occurs in the range  
220 to 280 min. The frequency of this modulation appears to be 
stable for a month or so, but no single period 
was consistently detected from one observing season to the next.
The occurrence of low states can be reconciled with a magnetic 
cataclysmic variable.  While the additional 
signal could point to an asynchronously rotating white dwarf, 
our data do not show this signal to be truly coherent.
We estimate the system luminosity by combining optical measurements
with the archival X-ray spectrum.  The implied mass 
accretion rate is orders of magnitudes below the 
predictions for the standard angular momentum loss above the period gap.  
The system may possibly belong to a largely undiscovered population 
of hibernating CVs. 


\end{abstract}

\section{Introduction}

Cataclysmic variable stars (CVs) are close binary systems in which a
low-mass secondary transfers mass onto a white dwarf; \citet{warn} wrote
an excellent monograph on CVs.

Because CVs are discovered by a variety of methods with idiosyncratic
and varying efficiency, their space density remains uncertain.
This is unfortunate, since an accurate knowledge of the space density would
serve as a strong constraint on evolutionary scenarios.

CV evolution is driven by the gradual loss of angular momentum.  As the
Roche lobe around the secondary star shrinks, matter is transferred to
the white dwarf, giving rise to the rich observed phenomenology.  The
mass transfer rate is observed to decline with declining orbital period
\citep{patterson84}, so there should be a preponderance of
short-period, slowly-evolving CVs, unless short-period CVs are somehow
destroyed (\citealt{patlate98} discusses these issues vividly).  In
addition, \citet{shara86} suggest that old novae `hibernate', that is,
render themselves faint and inconspicuous by fading into extended states
of little or no mass transfer between nova outbursts.  

The closer an object is, the easier it is to discover, so
representatives of these hard-to-find classes would be expected to show
up relatively nearby.  \citet{rbsid} summarize the properties of CVs
discovered in the ROSAT bright source (RBS) survey. 
The RBS was a program to identify optical counterparts of 
more than 2000 of the brightest X-ray sources, $CR > 0.2$ sec$^{-1}$, detected at
high latitudes ($|b| > 30\degr$) during the ROSAT all-sky survey.
The optical spectrum of one of these, since named V405 Peg 
\citep{namelist78}, shows a strong contribution from an M-type secondary.
This led \citet{rbsid} to suggest it was either a symbiotic-like
object or a CV; for the latter case they estimated a distance of only 30
pc.  \citet{kato-rbs1955} estimated a proper motion of $69 \pm 12$
milli-arcsec (mas) yr$^{-1}$ from catalog positions, which constrains
the distance to $< 300$ pc for transverse velocity $v_t < 100$ km
s$^{-1}$.  A 30 pc distance would make V405 Peg the nearest known CV,
and would suggest that a large space density of such objects remains
undetected.  We therefore observed this object to find its orbital
period, explore its nature, and constrain its distance.

\section{Observations and Analysis}
\label{s:obs}

{\it Spectroscopy.} We obtained spectra at the 2.4m Hiltner telescope at MDM
Observatory on Kitt Peak, using the modular spectrograph.  Most of the
spectra are from 2002 October, with sparser coverage during other observing
runs extending up to 2005 September (see Table 1).  The instrument
configuration and protocols were as described in \citet{t98}; briefly,
the setup yielded 2 \AA\ pixel$^{-1}$ from 4210 to 7500 \AA , with
severe vignetting towards the ends.  The wavelength calibration was
generally maintained to a few km s$^{-1}$, but for a few spectra the
$\lambda 5577$ night-sky line was used to correct calibration errors.
Flux standards were observed when the sky was reasonably clear.
Observations were extended to large hour angles to minimize ambiguities
in the orbital modulation's daily cycle count.  Exposure times were
mostly 300 s to resolve possible short periods.

{\it Direct imaging, astrometry, and color photometry.} Direct images
were obtained on eight observing runs, from 2002 October through
2005 November, using the
parallax protocols described by \citet{thorparallax}.  

On 2002 Oct. 25 UT we obtained a set of $UBVI$ images; we checked these
using $VI$ images obtained on two different photometric nights in 2003
June.  To derive transformations to standard magnitudes we observed
standard star fields from \citet{landolt}. The scatter of the standard
star magnitudes and colors was $\sim 0.03$ mag, and somewhat worse for
$U-B$.  Table \ref{t:stars} gives measurements of the brighter stars with their
estimated counting-statistics errors, together with positions derived
from a fit to numerous USNO-A2.0 stars \citep{mon96}.
Fig.~\ref{f:chart} shows an
$I$-band image of the field with $V$-magnitudes of selected stars.
V405 Peg proved to be the bluest object in the field in $U-B$ and $B-V$, 
but the {\it reddest} in $V-I$.

We selected 96 $I$-band images for the parallax reduction, for which we
used procedures described by 
\citet{thorparallax}.   The parallax relative to the reference-frame
stars, $\pi_{\rm rel}$, is 6.1 mas, with a formal error of 0.8 mas.
Based on the scatter of the field star parallaxes, we adopt a somewhat
more conservative uncertainty of 1.1 mas.  The colors and magnitudes of the 
reference stars indicate that the correction to absolute parallax is
near 1.1 mas, so our best estimate of $\pi_{\rm abs}$ is $7.2 \pm 
1.1$ mas.  The proper motion relative to the reference stars is 
$[-31, -50]$ mas yr$^{-1}$ in $X$ and $Y$ respectively, or 59 mas yr$^{-1}$ in
position angle 211 degrees.  The formal fitting error in the proper motion is
less than 1 mas yr$^{-1}$, but there are systematic uncertainties of 
at least a few mas yr$^{-1}$ in the correction to absolute proper 
motion (not attempted here); the reference stars in addition show a 
dispersion of several mas yr$^{-1}$ in each coordinate.
A second determination of the
proper motion alone, using archival Sky Survey images from the USNO Plate
Measuring Engine \footnote{available at
http://ftp.nofs.navy.mil/data/FchPix/} together with one of the CCD
images, produced a nearly identical result.

{\it Time resolved photometry.} 
Time resolved photometry was acquired on 48 nights 
in 2002, 2003, 2005 and 2006 using the 1.23~m reflector at Calar 
Alto Observatory, the 1~m Elisabeth telescope at SAAO, Sutherland, 
the IAC 80~cm telescope at Observatorio del Teide, Tenerife and 
the 70~cm telescope of the AIP in Babelsberg 
(see Table~\ref{t:obsphot} for a summary). 
To be more sensitive for the ellipsoidal variations of the 
secondary star all observations were taken with an $I$-filter,
but different filter systems had to be used; 
the Calar Alto and IAC observations were done with a
Johnson $I$-filter, while the SAAO natively uses the Kron-Cousins system.
The different band passes and effective wavelengths of the two filter
systems will introduce an additional uncertainty in the zero-point 
determination. \cite{Landolt83} and \cite{Bessell83} give
transformations for the $(R-I)$ color of both systems, which are
unfortunately not very well established for stars redder
than $(R-I)>1.2$ or later than spectral type M1. The transformations
indicate that for M-dwarfs independent of subtype the Johnson filter
will yield  $I$ magnitudes 0.06 mag brighter compared with the
Kron-Cousins system. We consequently have corrected the Calar Alto 
and IAC photometry by this amount.

We also performed an absolute calibration of the field stars at 1.23~m Calar 
Alto telescope with the Johnson $I$-filter, but using standards of
\cite{landolt} given in the Kron-Cousins system. 
The calibrated magnitudes of these agree within the errors with 
those obtained at the Hiltner telescope as predicted by the 
transformations cited above.
A detailed descriptions of the long-term variability of V405 Peg 
and period analysis will be given in Sect.~3.

{\it Spectrum and Flux Distribution.}  Fig.~\ref{f:spec}, the flux-calibrated
spectrum from 2002 October, shows one reason for the unusual colors 
-- the spectrum is a composite of an M dwarf and a much bluer component.
The average presented in Fig.~\ref{f:spec} excludes data obviously affected by
clouds, but the absolute flux is probably still underestimated by a
few tenths of a magnitude because of thin clouds and light losses at
the slit.  Table \ref{t:lines} gives measurements of the emission lines in the
mean spectrum.  

Fig.~\ref{f:lowhighspec} shows the mean spectra from 2002 October and December.
The spectrum varies dramatically.  In December the hot
component of the flux was much reduced and the emission lines were
much weaker, with the He I lines almost disappearing.  The Balmer
lines were much narrower than in October.  Evidently, the December
data represent a state of very low accretion.

We used the well-exposed 2002 October spectrum to estimate the
late-type contribution.  Archival spectra of M dwarfs taken with the
same instrumental setup were scaled and subtracted from the
object's average spectrum.  The archival spectra were all classified
by \citet{boeshaar}.  Types from M3 to M4+ gave acceptable
cancellation of the M-dwarf features over most of the spectral range,
but only M4 and M4+ matched the rapid rise of the continuum longward
of 7200 \AA , which is near the limit of the spectral coverage and
hence not as reliable as the middle.  The single M5 star library
spectrum did not give an acceptable subtraction.  The scale factors
giving good cancellation started near 0.6 of the combined flux (at
6500 \AA ) for type M3, and declined to 0.4 of the combined light at
type M4+ as the band strength increased.  The lower solid trace in
Fig.~\ref{f:spec} shows what remains after subtracting Gl 896B, classified 
M4+Ve by \citet{boeshaar}. \citealt{rbsid} classify the secondary as M3.
Considering all this we adopt  M3.75 $\pm 0.75$ subclasses for the
secondary.  This is similar to the spectral types of other CV
secondaries near this orbital period (4.27 hr, derived below;
\citealt{beuermann98} and \citealt{bk00} compile secondary spectral
types as a function of orbital period).

The M-dwarf contribution subtracted in Fig.~\ref{f:spec} has a synthetic $V =
17.66$, computed using the $V$ passband tabulated by \citet{bessell}.
The same procedure applied to the observed spectrum of V405 Peg yields $V =
16.07$, so for that particular decomposition, 0.23 of the $V$-band light
is from the secondary.  For the best M3 decompositions the corresponding
figures are $V = 17.24$ and 0.34, so a conservative estimate for the
secondary's fractional $V$-band contribution is $0.28 \pm 0.08$.  The
synthetic $V$ magnitude of the observed spectrum is $0.34$ magnitude fainter
than the $V$ magnitude measured with direct photometry on the same observing run.  
It is possible that the star varied by
this amount in the few-day interval between the direct and
spectroscopic observations,  but we believe it is likely that the cloud and slit
losses mentioned earlier account for the difference, and we
assume this is the case.  Accounting for this gives best estimates of
the secondary's $V$ magnitude of 17.32 and 16.90 for types M4+ and M3
respectively.  

Knowing the secondary's contribution, we can calculate the $V - I$ 
color of the `blue' component.  \citet{stauffer} tabulate accurate colors of M dwarfs
which also have spectral classifications by \citet{boeshaar}, and comparing these one
finds $V - I = 2.25$ for M3V stars and $V - I = 2.6$ for M4+V.  The
observed $V - I = 1.76$ and assumed contributions to the light then give
$V - I \sim +1.3$ for the `blue' component alone; the
calculation is insensitive to the secondary spectral type because the
secondary contribution gets weaker toward later (redder) spectral types.
The `blue' component is therefore surprisingly red in $V-I$.  

{\it Radial Velocities and Period.} We measured radial velocities of the
M-component by cross-correlating tone he line-rich 6000 - 6500 \AA\ spectral
region against an M-dwarf template, using {\it xcsao}
\citep{kurtzmink98}.  The template was created by taking spectra of
several M dwarfs with precise velocities measured by \citet{marcy},
shifting the spectra to zero velocity, and averaging them.  This
procedure should result in a velocity zero point accurate to  $\sim 5$
km s$^{-1}$.  The individual velocities of V405 Peg had typical
uncertainties of $\sim 15$ km s$^{-1}$ as estimated by {\it xcsao}.
There were 106 exposures that gave absorption velocities with estimated errors
less than 25 km s$^{-1}$; the remaining 17 velocities were not used.
For the H$\alpha$ emission lines, we measured velocities by convolving
the emission line profile with an antisymmetric function consisting of
positive and negative Gaussians $\pm 590$ km s$^{-1}$ from the line core
(see \citealt{sy}), which effectively gave a measurement of the line
wings.  The 116 spectra all yielded velocities, with typical
counting-statistics velocity uncertainties of a few km s$^{-1}$, but
with a less reliable zero point than the absorption velocities. 

The absorption-line velocity period search yields a period of 0.1776472(10)
d, or 4.26 hr.  There is no significant ambiguity in the cycle count over the
3-year span of the observations.  Because the 4.26-hour period is seen in the 
absorption lines, it is clearly the orbital period.  When the emission-line 
velocities are folded on this period, most of the data show a clear 
modulation with
little scatter, {\it except} for the ten velocities taken 2002 December,
when the star was in a much fainter state with narrower lines.  These
velocities did not show any modulation, but
instead scattered around the $\gamma$-velocity by $\sim 30$ km s$^{-1}$.  
When the 2002 December velocities are excluded, the remaining emission-line velocities
show a strong periodicity at the absorption period (as well as a few
other aliases allowed in by the exclusion of the December data).  The
combined best-fitting period is 0.1776469(7) d.  Table \ref{t:velocities}
gives sinusoidal fit parameters, and Fig.~\ref{f:velocities} shows folded 
velocities.  The emission
velocities are offset in phase from the absorption by 0.581(8) cycles,
compared to a value of 0.5 expected if the emission lines trace the 
white dwarf motion.  The absorption line velocities should accurately reflect the
phase of the M-dwarf motion, so the significant displacement 
of the emission velocities from their expected phase 
demonstrates that the emission velocities do {\it not} accurately trace 
the white dwarf motion.

With the binary period in hand, we prepared a single-trailed,
phase-averaged greyscale image of our 2002 October spectra.
Fig.~\ref{f:trail}
shows the region around H$\alpha$, which shows a nearly stationary,
narrow core with fainter wings, extending to $\pm 500$ km s$^{-1}$
from the line core.  The H$\alpha$ emission-line velocity measurements
appear to be dominated by the motion of these wings.  The HeI lines
(here represented by $\lambda 6678$) show a modulation similar to the
H$\alpha$ line wings.  The M-dwarf absorption features show clear
orbital velocity variation.  


\section{Time Series Photometry}
During the photometric monitoring 
V405~Peg exhibited two different brightness states which were 
accompanied by marked changes of its photometric variability.

{\it High state photometry 2002/2003}
In October/November 2002 and October 2003 the source was in a high state 
with an orbital averaged brightness of $I = 13.95$ and $I = 14.02$. 
At that time the light curves  exhibited a near
sinusoidal modulation with an amplitude of  $\sim 0.25$ mag,
with strong flickering activity superimposed.
A periodogram of these data sets (Fig.~\ref{f:period}) derived with 
the analysis of variance (AOV) method \citep{Schwarzenberg89} reveals a 
series of apparently significant periods in the range of 150 to 400 min, all of
which are one-day aliases of each other. 
The strongest of these aliases occur at 227.9 min, 270.0 min and
333.8 min, none of which coincides with the spectroscopic period. 
A separate AOV search for the October 2003 data confirms the reality 
of the independent periodic photometric signal, but does not 
resolve its daily cycle count ambiguity.    

The three different period solutions are illustrated in Fig.~\ref{f:light}
for the combined data of 2002. 
Visual inspection favors the two aliases closest to orbital period at 
227 and 270 min, which provide a slightly cleaner folding, especially
around the photometric minimum.
The two candidate periods would be
either 10 \% shorter or 5 \% longer than the binary period. 

{\it Low state photometry 2002/2003}
During December 2003 low state the source grew somewhat 
fainter, to $I=14\fm 15$, and exhibited a smooth, double-humped 
modulation without any flickering (Fig.~\ref{f:light}, lowest panel). 
This change also affects the periodogram (Fig.~\ref{f:period},
lower panel) where now a period close to the spectroscopic value of 255.8 min
(and its inevitable aliases) becomes the only notable signal. 
Very likely accretion has almost completely ceased at this epoch 
leaving the bare secondary star as the dominant source of light. 
The ellipsoidal variation with peak-to-peak amplitude of $\sim 0.1$ mag  
and two unequal minima are a clear signature of a tidally deformed companion. 
Contrary to the normal appearance of  ellipsoidal variation 
the secondary minimum is not symmetric to the primary, but displaced
from the timing of superior conjunction by 0.05 phase units. 
This small but significant effect may either indicate the influence
residual accretion or asymmetric distribution of the 
surface brightness of the secondary, e.g. by irradiation. 
The primary minimum is observed at HJD 2452979.3272.  This 
agrees within 0.01 cycle with the time predicted for inferior
conjunction of the M-star, using the absorption-line fit parameters
given in Table \ref{t:velocities}.

The long-term photometric coverage can be extended using 
differential photometry from the $I$-band parallax images.
For seven of the eight parallax observing runs,
the $I$-band magnitude was near 
$14.1$,
consistent with high state data from the time series photometry, but 
with significant variations of $\pm 0.1$ mag (while the comparison stars
were typically constant to within about $\pm 0.02$ mag).
In 2003 June the source was at a still higher level, near $I = 13\fm 7$.
Unfortunately we have no photometry close in time to the 
low state seen in the 2002 December spectroscopy (Fig.~\ref{f:lowhighspec}).

{\it High state photometry 2005}
A dense set of 20 nights of photometry was obtained in
September/October 2005. The source was in a high to intermediate state at
mean brightness of $I = 14\fm 0$. The periodogram (Fig.~\ref{f:aov_05})
computed from this data shows a complex pattern with two independent
periodicities. One signal is related the orbital period at 255.8 min, while the 
most prominent one appears at 251.8 min. 
The two periods are independent of each other, i.e. we detect no side-band 
periods between those two. The beat period on the 
other hand would be expected at 11.176 days. None of the periods previously
seen in the periodograms of the 2002 and 2003 are present in this data set.

In Fig.~\ref{f:lc05} the phase-folded light curves related to the two 
periodicities are displayed. The light curve folded on the orbital period 
shows a double-humped pattern similar to the low state of December 2003 and
is likely caused by ellipsoidal variation.
The dominant photometric signal at 251.8 min features a single humped 
light curve with a full amplitude of 0.05 mag. Thus the modulation 
was much weaker than during the observations of 2002/2003.

We also performed period searches 
using only sub-sets of the entire data. These experiments show that the
features in the periodograms are quite variable. 
For example, for the September data alone the additional periodicities appear
at 221 and 261 min, while no signal at 251.8 min, the peak of the combined 
data set, is detected. On the other hand, both signals at 221 and 261 min
do not show up in the combined data set.
Thus, the occurrence of the additional photometric periodicity is either 
time-dependent or strongly depends on the sampling of the data. 

{\it High state photometry 2006}
In order to
confirm the previously detected periods, a further set of 19
nights of photometry was obtained in September and November 2006. The 
source was again in a bright state, although the mean 
$I = 14.03$ was slightly fainter than in 2005.
In contrast to the periodogram from 2005 we can identify only one clear 
signal at the orbital period of 255.8 min. This detection is accompanied 
by a complicated pattern of alias periods, which includes the 
$\pm 1$~cycle/day and $\pm 0.5$~cycle/day ambiguities and a fine structure 
due to the two-week data gap in September 2006. We investigated the 
existence of further periodicities by subtracting the orbital 
modulation from the data. The periodogram of this cleaned data set 
(Fig.~\ref{f:period06})
shows tentative signals at 221.8, 262.4 and 278.4 min. 
Interestingly, two of those periods do coincide with the 
additional signals temporarily seen in September 2005. 
Nevertheless, our data do not specify a unique non-orbital 
periodicity that is coherent over different observing seasons.

\section{How Far Away is V405 Peg?}

We estimate the distance as follows.  First, we use secondary star's
contribution, in combination with the orbital period, to estimate a
distance spectrophotometrically.  Next, we use a Bayesian formalism
described by \citet{thorparallax} to combine this estimate with the
parallax and proper motion.  This gives a best possible distance. 

\citet{beuermann99} tabulate radii and other properties of a variety
of late-type stars.  From this one can derive the surface brightness
as a function of spectral type, which can be formulated conveniently
as $M_V(1 R_{\odot})$, the absolute $V$ magnitude of a 1-solar-radius
star of the same surface brightness.  Among the examples given in their
table, this quantity varies from a minimum of 8.61 to a maximum of 
9.97 in the range M3 -- M4.5.  From this, we adopt
$M_V(1 R_{\odot}) = 9.3 \pm 0.7$ for the M$3.75 \pm 0.75$ secondary
of V405 Peg.  

\citet{beuermann98} give a convenient formula for the Roche lobe
radius as a function of orbital period and secondary mass; at the
period of V405 Peg, this yields $R_2 / R_{\odot} = 0.616 f(q) (M_2 /
M_{\odot})^{1/3}$, where $f(q)$ is close to unity.  The secondary mass
$M_2$ can only be guessed at, but fortunately the radius depends only
weakly on $M_2$; as a
guide, the evolutionary models calculated by \citet{bk00} span a range
from 0.172 to 0.475 $M_{\odot}$ at $P_{\rm orb} = 4 $ hr, implying
$0.34 \le R_2 / R_{\odot} \le 0.48$.  Applying this to the surface
brightness range found above gives the secondary's absolute magnitude,
$M_V = 11.3 \pm 0.8$.   The spectral decomposition gave $V = 17.1 \pm
0.3$ for the secondary alone, which gives a distance modulus $m - M =
5.8 \pm 0.9$.  Combining these gives a best distance of 
$140 (+80,-50)$ pc \footnote{This
assumes the errors are uncorrelated.  In this case correlations
between the various derived quantities will tend to reduce the
uncertainty -- a later spectral type is intrinsically fainter, but
also contributes a lesser fraction of the light.} Note that this
computation does not assume a `normal' mass-radius relation for the
secondary -- it is based instead on the secondary's surface brightness 
and the requirement that it fill the Roche lobe.

Our Bayesian analysis of the parallax and proper motion follows
the procedure described by \citet{thorparallax}.  As noted earlier, 
we measure $\pi_{\rm abs} = 7.2 \pm 1.1$ mas.
We assume the same {\it a priori} distribution of space velocities
as \citet{thorparallax}.  Using the parallax and proper motion alone, the
Bayesian analysis gives $150(+28,-21)$ pc, consistent with the
spectrophotometric estimate.  Folding the spectrophotometric estimate
into the Bayesian analysis yields $149(+26,-20)$ pc.  The final
estimate is slightly greater than $1/\pi_{\rm abs}$ because a correction 
for the Lutz-Kelker bias increases distance estimates when the relative
parallax error is not small.  The rather small final
uncertainty reflects the concordance of all the information -- the
distance evidence is all nicely consistent.

\section{Orbit and dynamics}

{\it Binary Parameters.}  Combining the orbital period with the velocity 
amplitude of the secondary, $K_2 = 90 \pm 3$ km s$^{-1}$, yields a mass function $f(M) =
0.0134(14) M_{\odot}$.  Assuming broadly typical values of $M_1 = 0.7
M_{\odot}$ for the white dwarf and $M_2 = 0.3 M_{\odot}$ for the
secondary gives a binary inclination of 20 degrees (i.e., close to
face-on).  Decreasing $M_1$ to a very low $0.45 M_{\odot}$ brings the
inclination up only to 30 degrees,  so it is difficult to escape the
conclusion that the orbit is viewed close to face-on.  Because of the
phase offset between the emission and absorption velocities, we do
not believe further dynamical analysis is warranted.

\section{Broad-band spectral energy distribution}
The sky around V405 Peg happened to be scanned by GALEX, 2MASS, and SDSS 
photometrically, and SDSS also obtained a spectrum.
Fig.~\ref{f:sed1955} shows all those archival data together with our
low-state spectrum from Dec.~2002 and the ROSAT all sky survey spectrum.

The Sloan spectrum, obtained 2007 Aug 16, shows a rich emission-line
pattern, indicating another high-accretion state.
The overall flux level was comparable to that of our own high-state spectra. 
An M3.5 IR-optical template spectrum 
from Leggett's archive \citep{leggettetal00} scaled to fit the optical flux 
also accounts for the observed IR magnitudes from 2MASS.  There is 
therefore no evidence for an IR contribution in addition to the M dwarf;
in particular, there is no obvious cyclotron component.

An estimate to the temperature of the primary was obtained by matching 
the blue part of the optical low-state with a white dwarf template
assuming a DA spectral type with a standard mass of 0.6 M$_{\odot}$, 
a distance of 150\,pc,
and $T_{\rm eff} \leq 13000$\,K. A scaled model spectrum is included in 
Fig.~\ref{f:sed1955} but is regarded merely illustrative; 
the spectral type is not yet determined.  
GALEX photometry shows a pronounced excess over the extrapolated WD flux 
in both the FUV and NUV bands.
This could be due to stream emission, the heated 
photosphere below a hypothetical accretion spot or even due to cyclotron 
radiation from an accretion plasma in a high magnetic field.
An alternative way to estimate an upper limit temperature of the white dwarf
is derived on the requirement that an assumed white dwarf model atmosphere
scaled to the optical minimum spectrum shall not exceed the GALEX FUV
flux. This gives $T_{\rm eff,max} \simeq 17000$\,K and a massive white dwarf
of 1.1\,\msun\ for the assumed distance of 150\,pc. 

\citet{rbsid} used a two-component model to successfully reflect the
RASS-spectrum, consisting of a rather hot blackbody, 38 eV, and a thermal
component, likely from an accretion plasma. We wanted to derive more stringent
limits on the temperatures of the spectral components involved and thus 
retrieved the original photon data collected in the ROSAT all-sky survey. 
The X-ray spectrum of V405 Peg contains 101 photons. We found that the
observed spectrum can be fitted with an almost unabsorbed thermal
bremsstrahlung component only. The amount of interstellar absorption is small,
$N_{\rm H} \simeq 5\times 10^{19}$\,cm$^{-2}$, to make this
model work. Should the column density be higher, a soft blackbody-like component
is needed to reflect the spectrum. But there is no formal requirement from the
data to include such a component in the modelling. Inclusion of a soft
component leads to a large degeneracy between column density on the one hand
and the blackbody spectral parameters on the other hand. 

The one-component fit with the thermal component deteriorates for $k_{\rm B}T
< 3$\,keV, which we regard as lower limit temperature. Given the soft response
of ROSAT it is impossible to constrain the temperature at the high end. 
The bolometric flux of a 3\,keV thermal component is 
$F_{\rm th, bol} \simeq 3.5 \times 10^{-12}$\,erg cm$^{-2}$\,s$^{-1}$. 
Should the temperature be around
15\,keV, as one typically finds in accretion columns of magnetic
CVs, the bolometric flux becomes 
$F_{\rm th, bol} \simeq 8 \times 10^{-12}$\,erg cm$^{-2}$\,s$^{-1}$. 
The implied luminosity for a distance of 150\,pc and a geometry factor of
$2\pi$ is $L_{X} \simeq (4 - 10) \times 10^{30}$\,erg s$^{-1}$.

It is possible to hide a more luminous soft component in the soft X-ray/EUV
regime. The minimum temperature (maximum luminosity) of such a component which
still fits the X-ray spectrum and does not exceed the observed GALEX flux is
17\,eV, which gives 
$F_{\rm bb, bol} \simeq 4 \times 10^{-10}$\,erg cm$^{-2}$\,s$^{-1}$. 
At 25\,eV, which gives a fit to the X-ray spectra of
similar quality, the flux drops to 
$F_{\rm bb, bol} \simeq 3 \times 10^{-11}$\,erg cm$^{-2}$\,s$^{-1}$, still
higher than the flux in the thermal component. However, the lack of 
HeII emission lines and the fact that the X-ray spectrum is well explained
with just a thermal component makes it unlikely that a luminous soft X-ray
component exists in V405 Peg.

The difference between optical high- and low-state spectra give an integrated
flux of $F_{\rm opt} \simeq 2 \times 10^{-12}$\,erg cm$^{-2}$\,s$^{-1}$. 
A bolometric correction factor could be of order $2-4$, but is almost
unconstrained. Hence, the accretion-induced optical flux is likely smaller
than the X-ray flux although a broader wavelength coverage in the optical and
the X-ray ranges is necessary to reach more definite conclusions.

The implied accretion luminosity of optical and X-ray components is poorly
constrained, but likely in the range $4 \times 10^{30} - 5 \times
10^{31}$\,erg\,s$^{-1}$ (the implied mass accretion rate $\dot{M} =
(R_{\rm wd}/GM_{\rm wd}) L_{\rm acc}$ is 
$7\times 10^{-13} - 8 \times 10^{-12}$\,M$_{\odot}$\,yr assuming $M_{\rm wd} =
0.6$\,M$\odot$ and $R_{\rm wd} = 8.5 \times 10^{10}$\,cm).
This is too low by a factor $\sim$100 at the given binary period for a
typical polar system.  However, it is also $\sim 100$ times greater
than found for the magnetic pre-cataclysmic systems discovered in the HQS or
the SDSS \citep{Reimers99,schmidtlarp,vogelwx}. 

\section{What kind of CV is V405 Peg?}
\label{s:whatisit}
Our observations show that V405 Peg is a low-accretion-rate CV
with a period well longward of the 2 to 3 hour `period gap'.
The residual accretion observed at several epochs 
proves the existence of mass transfer,
but its rate is several orders of 
magnitudes lower than the value of 10$^{-9}$ M$_{\odot}$~yr$^{-1}$ 
empirically found for long period CVs and implied by magnetic braking 
\citep{patterson84}.  
While there is the small probability that V405 Peg is a previously detached 
white dwarf/M-dwarf binary just establishing contact, it is more likely
that it has been a normal CV in the past temporarily accreting 
at a low rate.
One possiblity is that it is in an extended state of hibernation
\citep{shara86}. Such long-term cyclic changes of the mass transfer rate 
are believed to be the response of a CV to the effects of a 
nova explosion and might be the underlying cause for the large observed 
spread in mass transfer rates of CVs at a given orbital period.
However, broader acceptance of this hypothesis still suffers from the 
difficulty of confirming a decline in the mass-transfer rate in any of
the recorded post novae.
The discovery of V405 Peg, together with the recent identification 
of a number of DA-dM binaries  with almost-attached secondaries
(\citealt{ODonoghue03}, \citealt{Kawka02}, \citealt{Gaensicke04}) in the 
period range preferentially occupied by old 
novae \citep{Warner02} is a first indication that a substantial number 
of hibernating CVs may exist.
In contrast to the DA-dM binaries which show no obvious signs of accretion,
V405 Peg would be the first candidate system occasionally accreting at a 
low rate. 

An alternate hypothesis would be that V405 Peg is a CV of the VY Sculptoris
sub-class currently stalled in an extended low state\footnote{Even in its 
high state, V405 Peg is much less luminous than typical VY Scl stars in their 
high state; see, e.g., Table 1 of \citealt{Townsley09}.}.  In these
high-$\dot M$, nova-like CVs, which preferentially cluster at orbital
periods between 3 and 4 hr, mass transfer is temporarily interrupted 
at random intervals.  However, the white dwarfs in these 
systems are very hot with $T_{\rm eff}$ in the range of 40 to 50 kK 
\citep{Hamilton08}, in contrast to the upper limit of 17 kK for V405 Peg 
set by the GALEX fluxes.  The timescale for cooling the heated envelope 
of the white dwarf to an effective temperature of 17 kK is $5\times 10^4$~yr 
\citep{Townsley09}, which provides a rough estimate for the time since its
last  nova erruption or the period spent in a VY Scl type low-state.  Because
of the long implied cooling time, we conclude that V405 Peg is unlikely
to be a long-dormant VY Scl star.

V405 Peg exhibits two features that suggest it is a magnetic CV, namely
(1) frequent changes between episodes of residual accretion and complete 
off states, and (2) a photometric signal unrelated to the orbital period.
The latter could be interpreted as a signal of an asynchronously rotating
white dwarf. However, the hallmark of such a system would be a
truly stable period, which our extensive photometric data fail to show.
Either the accretion process is highly unstable in V405 Peg, which would
be unusual for such a low accretion rate system, or the sampling of 
our data is still not sufficient.  Also, most magnetic cataclysmics
show stronger HeII $\lambda$ 4686 than seen here, though in low
states all the emission lines can be much reduced (e.g., \citealt{mason07}).

The spectral energy distribution unfortunately does not help to constrain the
likely type of CV. If the object is strongly magnetic, i.e.~of a polar type,
what would be the field strength?  In which part of the spectrum would the
cyclotron spectrum be located?  The absence of a unique spin period equaling
the orbital period  argues against the polar interpretation. On the other
hand, the object also  does not fit comfortably as an intermediate polar.  Why
should it be asynchronous given its low accretion rate?  

The low luminosity suggests that V405 Peg might be related to the low 
accretion rate polar (LARP) systems found in the Sloan Digital Sky Survey 
sample \citep{schmidtlarp,vogelwx}.  These appear to be pre-polars that 
underfill their Roche lobes and are powered by wind accretion via a magnetic
siphon from the late type secondary.  However, there are
significant dissimilarities between V405 Peg and the LARPs:
(1) V405 Peg does not show the the narrow cyclotron features seen in LARPs.
(2) X-ray luminosity and accretion 
rate in V405 Peg are at least one order of magnitude higher compared to 
the pre-polars \citep{schmidtlarp,vogelwx}.
(3) The transitions between residual accretion and complete 
cessation of the mass transfer are morphologically
similar to the high/low state behavior of AM Her stars, but have
not yet been observed in LARPs.
(4) The optical modulation from the cyclotron spots in LARPs tends
to be very smooth, probably as a consequence of wind accretion. 
The high-state light curves of V405 Peg, on the other hand, show strong 
flickering, which arises naturally in disk or an accretion stream,
either of which would indicate Roche-lobe overflow.

Many disk-accreting systems in this period range show so-called {\it permanent
superhumps}, photometric modulations at frequencies somewhat different
from the orbital frequency (see, e.g., \citealt{patt02} and references
therein).  However, all these systems are much more luminous than V405
Peg.  It appears that this phenomenon arises from disk precession, and
the instability which drives the precession requires the disk to occupy
a substantial portion of the Roche lobe radius; this in turn requires
the disk to be persistently bright.  It is therefore very unlikely that
the non-orbital photometric modulation is a superhump.

If V405 Peg is not a magnetic system, then it should have an accretion
disk.  At low inclination, emission lines are generally single-peaked,
as observed.  However, the disk hypothesis does not fit comfortably with
the strong variability of the emission-line spectrum seen in 
Fig.~3; the line spectra
of dwarf novae at minimum light tend to be relatively steady.  Also, if
V405 Peg were a disk system, it would be expected to undergo dwarf nova
eruptions from time to time, but outbursts have not been recorded.
Using the \citet{warn87} relation between  $P_{\rm orb}$ and outburst
absolute magnitude $M_{V(\rm max)}$, together with our period, distance,
and inclination, we estimate that dwarf nova
outbursts from this object would reach $V = 9.5$, so if outbursts occur
they would escape notice only if they were very infrequent.
Some dwarf novae, such as WZ Sge, do outburst only every few decades, and
the high-proper-motion star GD552 \citep{hessmangd552} resembles a dwarf
nova spectroscopically but has never been observed to outburst, implying
a still longer outburst interval.  These seldom-outbursting systems have
orbital periods $< 2$ hr, however, much shorter than V405 Peg.
\citet{tappertcw1045} suggest that LY UMa (= CW 1045+525), with 
$P_{\rm orb} = 0.271$ d, may be a long-period example of a dwarf nova which seldom
outbursts, but there is no clear evidence for a disk in that system
either. 


In sum, we are not able to unambiguously classify V405 Peg as a known
type of CV -- it is not clear which cage it should occupy in the CV
`zoo' -- but the magnetic classification fits most comfortably.  The
cause of the photometric modulation remains an open question.

\section{Conclusion}

We have shown that V405 Peg is relatively nearby for a CV, but not
extraordinarily so.  The prominence of its secondary star shows that the
accretion rate is very low.  It is difficult to classify the system
unambiguously, but it is most likely a magnetic CV.  The photometric
modulation we find remains unexplained.

\acknowledgements 
JRT gratefully acknowledges support from the U.S. National Science
Foundation, though grants AST-9987334, AST-0307413, and AST-0708810.  
S\'ebastien L\'epine and Bill Fenton took some of the MDM observations.
This research has made use of the USNOFS Image and Catalogue Archive
operated by the United States Naval Observatory, Flagstaff Station.
RS, A. Staude, JV, MK, and ANGM are supported by the Deut\-sches Zentrum f\"ur
Luft- und Raumfahrt (DLR) GmbH under contracts No. FKZ  
\mbox{50 OR 0206} and \mbox{50 OR 0404}.  JK was supported by the
DFG under contract Schw536/23-1.  We thank the anonymous referee for
numerous constructive suggestions that led to a significantly stronger
paper.


\clearpage
\begin{deluxetable}{lcrcc}
\label{t:obsmdm}
\tabletypesize{\small}
\tablewidth{0pt}
\tablecolumns{6}
\tablecaption{Journal of MDM Observations}
\tablehead{
\colhead{UT Date} &
\colhead{Instrument\tablenotemark{a}} &
\colhead{$N$} &
\colhead{HA start} &
\colhead{HA end} 
}
\startdata
2002 Oct 24 & D &  9 & $ -1:23$ & $ +1:28$ \\
2002 Oct 25 & D & 14 & $ +0:21$ & $ +0:58$ \\
2002 Oct 26 & S & 49 & $ -2:00$ & $ +4:57$ \\ 
2002 Oct 27 & S & 16 & $ +0:23$ & $ +1:54$ \\ 
2002 Oct 29 & S & 19 & $ -2:03$ & $ -0:12$ \\ 
2002 Oct 30 & S &  7 & $ -0:39$ & $ -0:03$ \\ 
2002 Oct 31 & S &  9 & $ +0:00$ & $ +0:47$ \\ 
2002 Dec 12 & S &  3 & $ +0:42$ & $ +3:06$ \\ 
2002 Dec 13 & S &  2 & $ +2:51$ & $ +2:57$ \\ 
2002 Dec 14 & S &  4 & $ +2:23$ & $ +4:15$ \\ 
2002 Dec 15 & S &  1 & $ +1:20$ & $ +1:20$ \\ 
2003 Feb 01 & S &  1 & $ +4:30$ & $ +4:30$ \\ 
2003 Feb 02 & S &  2 & $ +4:46$ & $ +4:55$ \\ 
2003 Jun 17 & D &  7 & $ -2:00$ & $ -1:42$ \\
2003 Jun 19 & D & 11 & $ -1:58$ & $ -1:27$ \\
2003 Jun 22 & S &  1 & $ -1:37$ & $ -1:37$ \\ 
2003 Jun 25 & S &  1 & $ -4:18$ & $ -4:18$ \\ 
2003 Oct 09 & D & 11 & $ +0:27$ & $ +0:56$ \\
2003 Oct 12 & S &  1 & $ +1:55$ & $ +1:55$ \\ 
2004 Jan 13 & S &  1 & $ +4:00$ & $ +4:00$ \\ 
2004 Jan 16 & S &  1 & $ +2:56$ & $ +2:56$ \\ 
2004 Jun 20 & D &  6 & $ -2:07$ & $ -1:51$ \\
2004 Jun 21 & D &  6 & $ -1:53$ & $ -1:39$ \\
2004 Jun 25 & S &  1 & $ -1:07$ & $ -1:07$ \\ 
2004 Nov 12 & D &  5 & $ +0:44$ & $ +1:02$ \\
2005 Jun 26 & D &  3 & $ -0:54$ & $ -0:48$ \\
2005 Jun 27 & D &  5 & $ -1:45$ & $ -1:29$ \\
2005 Jun 30 & S &  1 & $ -2:12$ & $ -2:12$ \\ 
2005 Sep 08 & S &  3 & $ -1:47$ & $ -1:29$ \\ 
2005 Sep 13 & D & 10 & $ +0:01$ & $ +0:29$ \\
2005 Sep 14 & D &  6 & $ +0:13$ & $ +0:30$ \\
2005 Sep 15 & D &  7 & $ -0:02$ & $ +0:17$ \\
2005 Nov 18 & D & 12 & $ +0:51$ & $ +1:23$ \\
\enddata
\tablenotetext{a}{D and S refer to direct and
spectroscopic observations, respectively.}
\end{deluxetable}

\clearpage
\begin{deluxetable}{llrrr}
\tablewidth{0pt}
\tablecolumns{5}
\tablecaption{Journal of $I$-band time series}
\tablehead{
\colhead{UT Date} &
\colhead{Site} &
\colhead{Exposure (sec)} &
\colhead{$N$} &
\colhead{Duration (h)} 
}
\startdata
2002 Oct 22 & CA 1.23m &  60 & 223 & 4.38  \\
2002 Oct 23 & CA 1.23m &  60 & 81  & 1.58  \\
2002 Oct 25 & CA 1.23m &  60 & 136 & 2.68 \\ 
2002 Nov 18 & CA 1.23m &  60 & 114 & 3.95 \\ 
2002 Nov 25 & CA 1.23m &  60 & 64  & 1.24 \\ 
2003 Oct 26 & SAAO  1m &  60 & 256 & 4.37 \\ 
2003 Oct 27 & SAAO  1m &  60 & 273 & 4.60 \\ 
2003 Dec 03 & IAC 80cm & 150 &  36 & 1.83 \\ 
2003 Dec 05 & IAC 80cm &  90 & 159 & 4.74 \\ 
2003 Dec 07 & IAC 80cm &  50 & 148 & 5.02 \\ 
2003 Dec 08 & IAC 80cm &  50 & 111 & 4.16 \\ 
2005 Aug 30 & AIP 70cm & 120 &  224 & 7.93 \\
2005 Aug 31 & AIP 70cm & 120 &  191 & 6.63 \\
2005 Sep 05 & AIP 70cm & 120 &  140 & 7.71 \\
2005 Sep 01 & AIP 70cm & 120 &  221 & 7.88 \\
2005 Sep 06 & AIP 70cm & 120 &  490 & 8.66 \\
2005 Sep 07 & AIP 70cm & 120 &  234 & 8.42 \\
2005 Sep 08 & AIP 70cm & 120 &  240 & 8.18 \\
2005 Sep 19 & AIP 70cm & 120 & 241 & 8.73 \\ 
2005 Oct 04 & IAC 80cm & 40  & 385 & 5.53 \\
2005 Oct 05 & IAC 80cm & 40  & 469 & 7.30 \\
2005 Oct 05 & AIP 70cm & 120 &  78 & 2.73 \\
2005 Oct 08 & IAC 80cm & 40  & 297 & 4.08 \\
2005 Oct 09 & IAC 80cm & 40  & 356 & 4.90 \\
2005 Oct 09 & AIP 70cm & 120 & 101 & 4.07 \\
2005 Oct 10 & IAC 80cm & 40  & 385 & 5.24 \\
2005 Oct 11 & AIP 70cm & 120 & 417 & 7.51 \\
2005 Oct 13 & IAC 80cm & 40  & 563 & 7.75 \\
2005 Oct 13 & AIP 70cm & 120 &  47 & 1.61 \\
2005 Oct 19 & AIP 70cm & 120 &  90 & 3.26 \\
2005 Oct 27 & AIP 70cm & 120 & 126 & 4.51 \\
2006 Sep 01 & IAC 80cm & 60 & 388 & 7.44 \\
2006 Sep 02 & IAC 80cm & 60 & 146 & 3.59 \\
2006 Sep 04 & IAC 80cm & 60 & 374 & 7.65 \\
2006 Sep 05 & IAC 80cm & 60 & 282 & 5.39 \\
2006 Sep 06 & IAC 80cm & 60 & 225 & 6.16 \\
2006 Sep 08 & IAC 80cm & 60 & 135 & 2.60 \\
2006 Sep 09 & IAC 80cm & 60 & 109 & 2.08 \\
2006 Sep 11 & AIP 70cm & 90 & 328 & 8.81 \\
2006 Sep 20 & AIP 70cm & 90 & 340 & 9.27 \\
2006 Sep 21 & AIP 70cm & 90 & 791 & 6.87 \\
2006 Sep 22 & AIP 70cm & 90 & 347 & 9.50 \\
2006 Sep 24 & AIP 70cm & 90 & 261 & 7.24 \\
2006 Nov 21 & IAC 80cm & 60 & 85 & 1.68 \\
2006 Nov 22 & IAC 80cm & 90 & 159 & 4.37 \\
2006 Nov 23 & IAC 80cm & 90 & 99 & 2.67 \\
2006 Nov 25 & IAC 80cm & 90 & 199 & 5.42 \\
2006 Nov 26 & IAC 80cm & 90 & 187 & 5.09 \\
2006 Nov 27 & IAC 80cm & 90 & 190 & 5.16 \\
2006 Nov 28 & IAC 80cm & 90 & 104 & 2.83 \\
\enddata
\label{t:obsphot}
\end{deluxetable}

\clearpage
\begin{deluxetable}{llllll}
\tablewidth{0pt}
\tablecolumns{6}
\tablecaption{Positions and Magnitudes}
\tablehead{
\colhead{$\alpha_{\rm ICRS}$} &
\colhead{$\delta_{\rm ICRS}$} &
\colhead{$V$} &
\colhead{$U-B$} &
\colhead{$B-V$} & 
\colhead{$V-I$}  
}
\startdata
\cutinhead{\it V405 Peg: \hfill}
 23:09:49.15 & +21:35:17.2  &   15.727(5)   &   $-$1.096(5)   &   $-$0.193(6)   &   1.759(9)   \\ 
\cutinhead{\it Field stars: \hfill}
 23:09:35.27 & +21:38:51.1  &   16.352(10)   &   0.161(15)   &   0.812(12)   &   0.920(19)   \\ 
 23:09:36.76 & +21:37:16.5  &   16.469(6)   &   0.653(19)   &   0.996(8)   &   1.144(8)   \\ 
 23:09:42.12 & +21:36:16.7  &   16.301(6)   &   $-$0.038(13)   &   0.576(7)   &   0.787(9)   \\ 
 23:09:44.82 & +21:35:37.5  &   17.220(7)   &   0.30(2)   &   0.825(10)   &   0.938(12)   \\ 
 23:09:45.10 & +21:32:06.2  &   17.254(7)   &   1.08(4)   &   1.101(11)   &   1.261(13)   \\ 
 23:09:46.47 & +21:38:01.5  &   17.478(8)   &   0.91(4)   &   1.058(11)   &   1.193(10)   \\ 
 23:09:48.83 & +21:31:50.6  &   17.178(8)   &   0.54(2)   &   0.919(11)   &   1.084(15)   \\ 
 23:09:53.93 & +21:32:03.8  &   16.920(8)   &   0.120(17)   &   0.725(11)   &   0.862(15)   \\ 
 23:09:56.02 & +21:32:43.1  &   16.803(6)   &   0.173(16)   &   0.737(8)   &   0.874(12)   \\ 
 23:09:56.60 & +21:35:27.9  &   17.717(8)   &   0.65(4)   &   0.987(13)   &   1.200(11)   \\ 
 23:09:56.79 & +21:36:18.9  &   17.480(7)   &   1.35(8)   &   1.325(12)   &   1.447(10)   \\ 
 23:09:57.34 & +21:33:57.0  &   14.486(2)   &   0.508(7)   &   0.907(4)   &   0.969(5)   \\ 
 23:09:57.40 & +21:35:17.1  &   17.656(7)   &   0.22(2)   &   0.767(11)   &   0.848(11)   \\ 
 23:09:57.61 & +21:35:41.4  &   17.897(8)   &   0.91(6)   &   1.069(14)   &   1.144(12)   \\ 
 23:09:58.90 & +21:38:16.2  &   17.497(8)   &   1.34(11)   &   1.470(14)   &   1.625(11)   \\ 
 23:10:00.42 & +21:33:10.6  &   16.924(8)   &   0.114(17)   &   0.737(10)   &   0.896(14)   \\ 
 23:10:03.70 & +21:35:28.5  &   17.026(17)   &   1.11(5)   &   1.34(2)   &   1.74(4)   \\ 
 23:10:05.13 & +21:36:06.1  &   14.838(6)   &   0.198(8)   &   0.831(8)   &   0.925(9)   \\ 
\enddata
\label{t:stars}
\end{deluxetable}

\begin{deluxetable}{lrcc}
\tablewidth{0pt}
\tablecolumns{4}
\tablecaption{Emission Features}
\tablehead{
\colhead{Feature} & 
\colhead{E.W.\tablenotemark{a}} & 
\colhead{Flux\tablenotemark{b}}  & 
\colhead{FWHM \tablenotemark{c}} \\
 & 
\colhead{(\AA )} & 
\colhead{} & 
\colhead{(\AA)} 
}
\startdata
\sidehead{2002 October} 
         H$\gamma$ & $ 83$ & $173$ & 15 \\ 
HeI $\lambda 4388$ & $  3$ & $ 6$ & 14 \\ 
HeI $\lambda 4471$ & $ 24$ & $41$ & 17 \\ 
HeII $\lambda 4686$ & $  5$ & $ 7$ & 15 \\ 
HeI $\lambda 4713$  & $  2$ & $ 4$ & 20 \\ 
         H$\beta$ & $116$ & $178$ & 16 \\ 
HeI $\lambda 4921$ & $ 14$ & $21$ & 17 \\ 
HeI $\lambda 5015$ & $ 10$ & $15$ & 14 \\ 
 Fe $\lambda 5169$ & $  8$ & $11$ & 15 \\ 
HeI $\lambda 5876$ & $ 33$ & $44$ & 18 \\ 
        H$\alpha$ & $108$ & $187$ & 17 \\ 
HeI $\lambda 6678$ & $ 14$ & $21$ & 19 \\ 
\sidehead{2002 December} 
H$\beta$ & $ 33$ & $10$ & 11 \\ 
H$\alpha$ & $ 22$ & $26$ &  9 \\ 
\enddata
\tablenotetext{a}{Emission equivalent widths are counted as positive.}
\tablenotetext{b}{Absolute line fluxes in units of $10^{-15}$ erg
cm$^{-2}$ s$^{-1}$.  These are are uncertain by about
30 per cent, but relative fluxes of strong lines
should be accurate to $\sim 10$ per cent.} 
\tablenotetext{c}{From Gaussian fits.}
\label{t:lines}
\end{deluxetable}

\clearpage
\begin{deluxetable}{lrrrrcc}
\footnotesize
\tablewidth{0pt}
\tablecaption{Fits to Radial Velocities}
\tablehead{
\colhead{Data set} & 
\colhead{$T_0$\tablenotemark{a}} & 
\colhead{$P$} &
\colhead{$K$} & 
\colhead{$\gamma$} & 
$N$ & 
\colhead{$\sigma$} \\ 
\colhead{} & 
\colhead{} &
\colhead{(d)} & 
\colhead{(km s$^{-1}$)} &
\colhead{(km s$^{-1}$)} & 
\colhead{} &
\colhead{(km s$^{-1}$)} \\
}
\startdata
Absorption & 52623.6767(10) &  0.1776472(10) &  92(3) & $-23(2)$ & 106 &  16 \\
H$\alpha$ emission & 52671.5669(9) & 0.1776467(9) &  73(2) & $-33(2)$ & 113 &  12 \\
Combined: & \nodata & 0.1776469(7) & \nodata & \nodata & \nodata & \nodata \\
\enddata
\tablenotetext{a}{Blue-to-red crossing, HJD $- 2450000$.}
\label{t:velocities}
\end{deluxetable}

\clearpage

\begin{figure}
\caption{Finding chart for V405 Peg 
(ICRS coordinates 
$\alpha = 23^{\rm h} 09^{\rm m} 49^{\rm s}.15, 
\delta = +21^{\circ} 35' 17''.2$)
from an average of several 60-s
I-band exposures taken with the MDM 2.4m.  Some of the field stars are labeled
with their $V$ magnitudes.  The scale and orientation are indicated.
}
\label{f:chart}
\end{figure}

\begin{figure}
\plotone{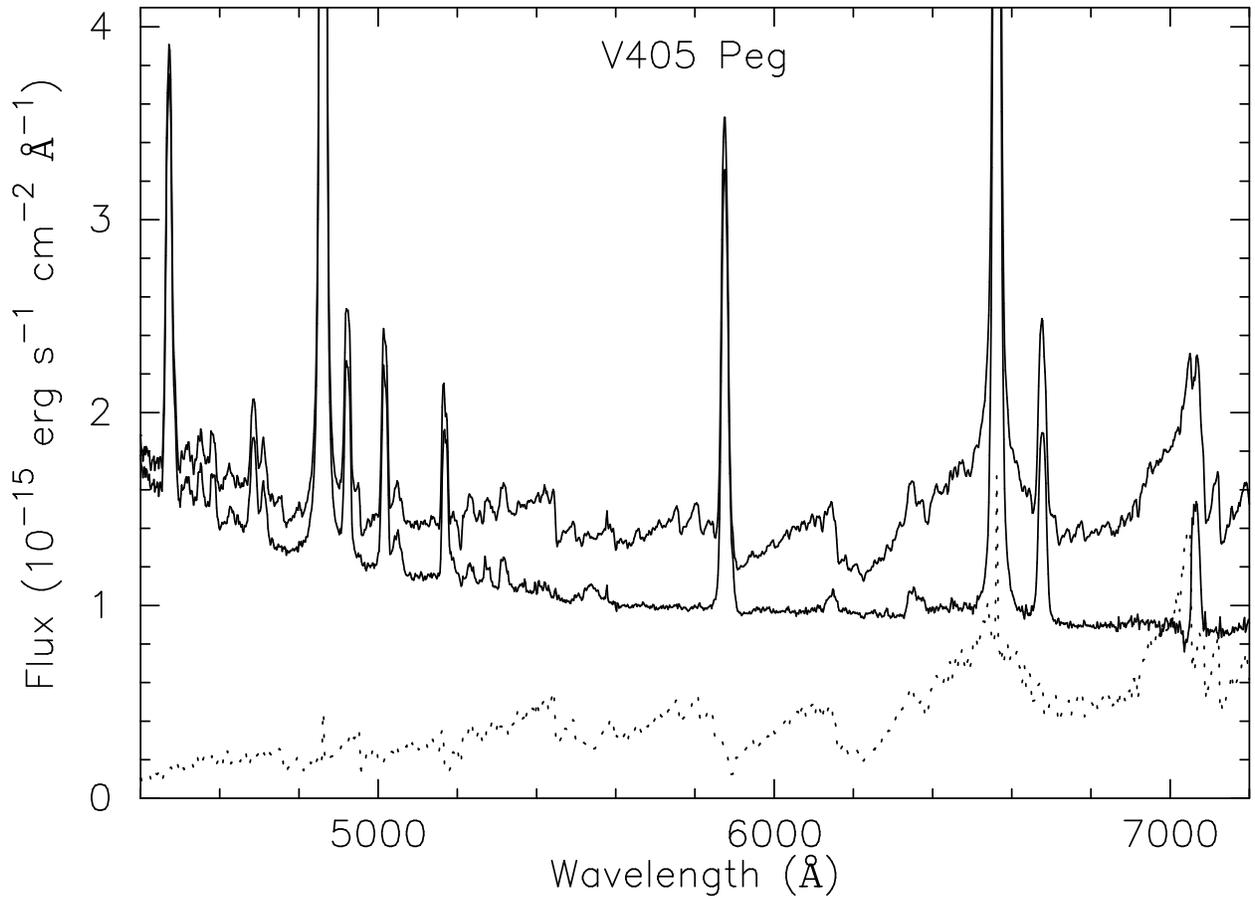}
\caption{{\it Upper trace}: Mean flux-calibrated spectrum from 
2002 October.  {\it Lower trace}: The same, minus a spectrum of the 
M3 dwarf Gl896B scaled
to $10^{-15}$ erg cm$^{-2}$ s$^{-1}$ \AA$^{-1}$ at 
$\lambda = 6500$ \AA .  {\it Dotted line}: The scaled M-dwarf
spectrum.
}
\label{f:spec}
\end{figure}

\clearpage

\begin{figure}
\plotone{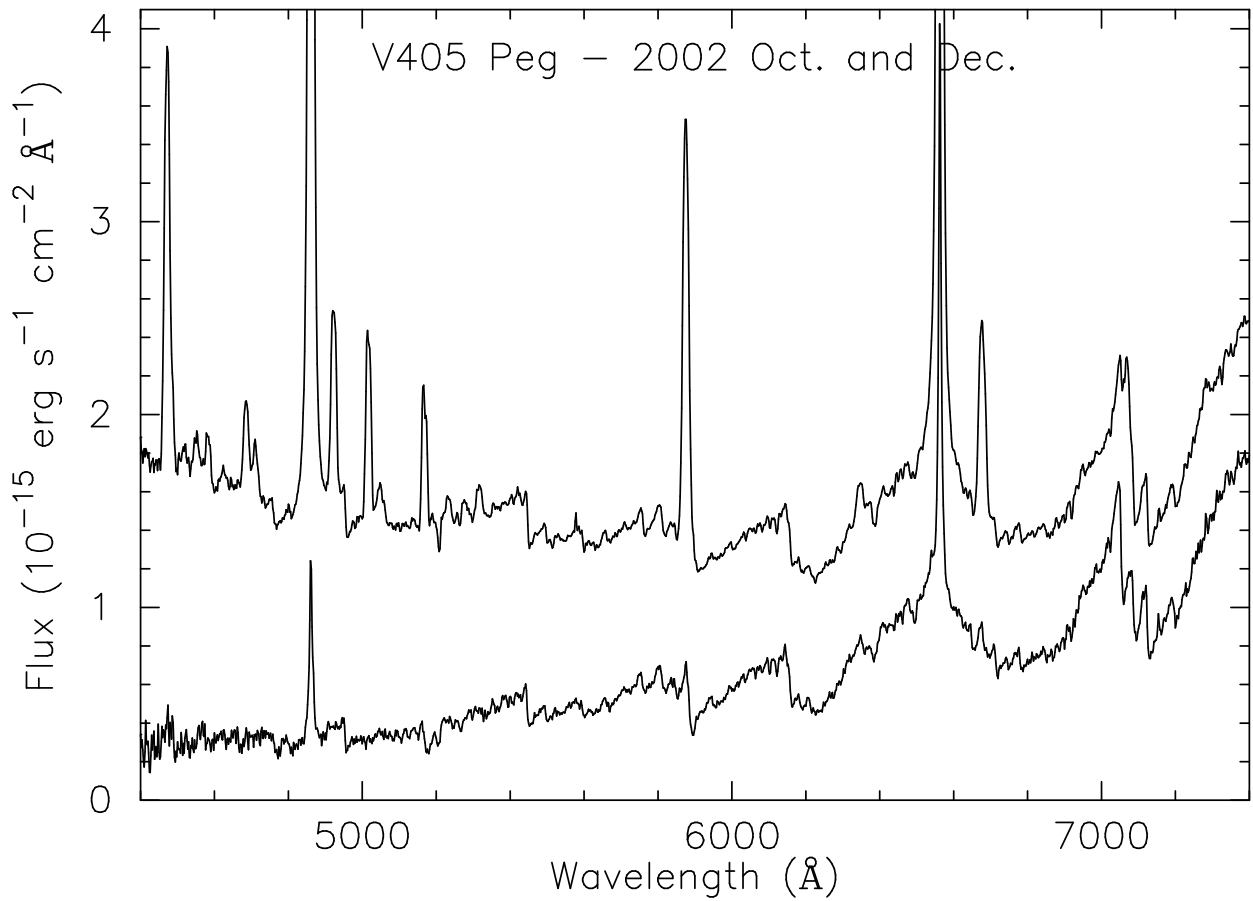}
\caption{(Upper trace).  Mean spectrum from 2002 October.  (Lower
trace).  Mean spectrum from 2002 December, showing markedly weaker
emission lines and continuum.
}
\label{f:lowhighspec}
\end{figure}

\clearpage

\begin{figure}
\plotone{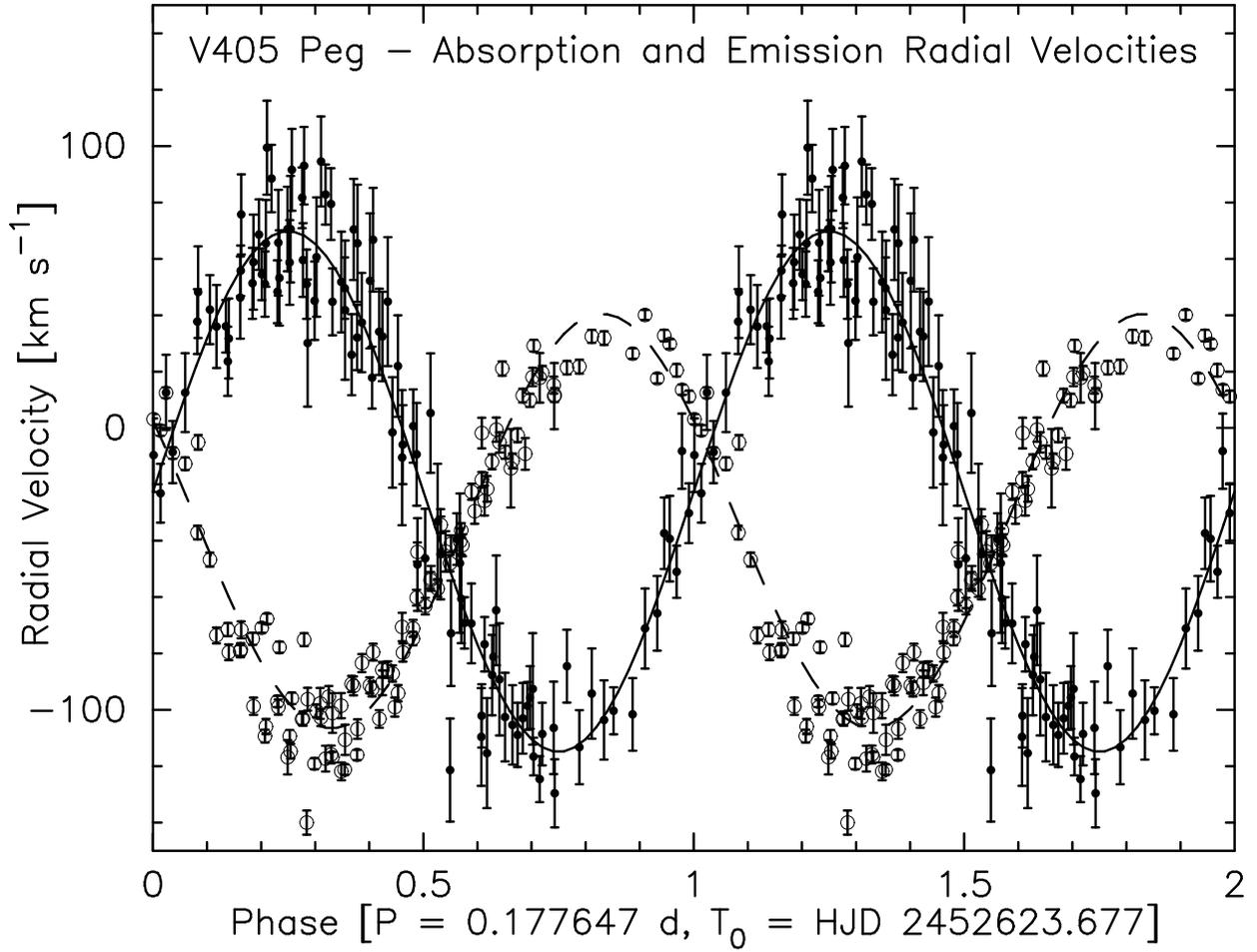}
\caption{Absorption (solid round dots) and H$\alpha$
emission (open squares) velocities folded on the 
binary period.  The best-fitting sinusoids are plotted.
}
\label{f:velocities}
\end{figure}

\clearpage

\begin{figure}
\caption{The H$\alpha$ and HeI $\lambda 6678$ region of spectra from 
2002 October, rectified and shown as greyscale 
against phase.  The greyscale is negative (dark = emission).  In the 
lower panel the scale is chosen to emphasize the cores of emission features
features, and in the upper it is chosen to show the absorption lines.
All data are shown twice for continuity.
\label{f:trail}
}
\end{figure}

\clearpage
\begin{figure}
\epsscale{0.8}
\bigskip\bigskip\bigskip\bigskip
\caption{$I$-band light curves obtained in October/November 2002 (upper
four panels), in October 2003 (5th panel) and December 2003 (bottom
panel). The 2002 data are plotted as the function of the three 
photometric aliases and the spectroscopic 
period,  while the high state light curve of October 2003 are folded 
with only one of possible aliases.
For the  low state in December 2003 the orbital ephemeris given in
Sect.~\ref{s:obs} is used.  
All data are shown twice for continuity.
The error bars to the right indicate the average error for each 
observing run.
}
\label{f:light}
\end{figure}

\clearpage
\begin{figure}
\epsscale{1.0}
\caption{Three different AoV-periodograms computed from the $I$-band data
of October-November 2002, October and December 2003.
}
\label{f:period}
\end{figure}

\clearpage
\begin{figure}
\caption{AoV-periodogram computed from the data set
of September-October 2005.
}
\label{f:aov_05}
\end{figure}

\begin{figure}
\caption{AoV-periodogram computed from the data set
of September-November 2006.
}
\label{f:period06}
\end{figure}

\clearpage
\begin{figure}
\epsscale{0.7}
\caption{Phase-folded light curve of the September/October 2005 data set
folded using the orbital period (upper panel) and the dominant photometric
periodicity at 251.8 min (lower panel).
}
\label{f:lc05}
\end{figure}

\clearpage
\begin{figure}[t]
\resizebox{0.5\width}{!}{\includegraphics[clip=true]{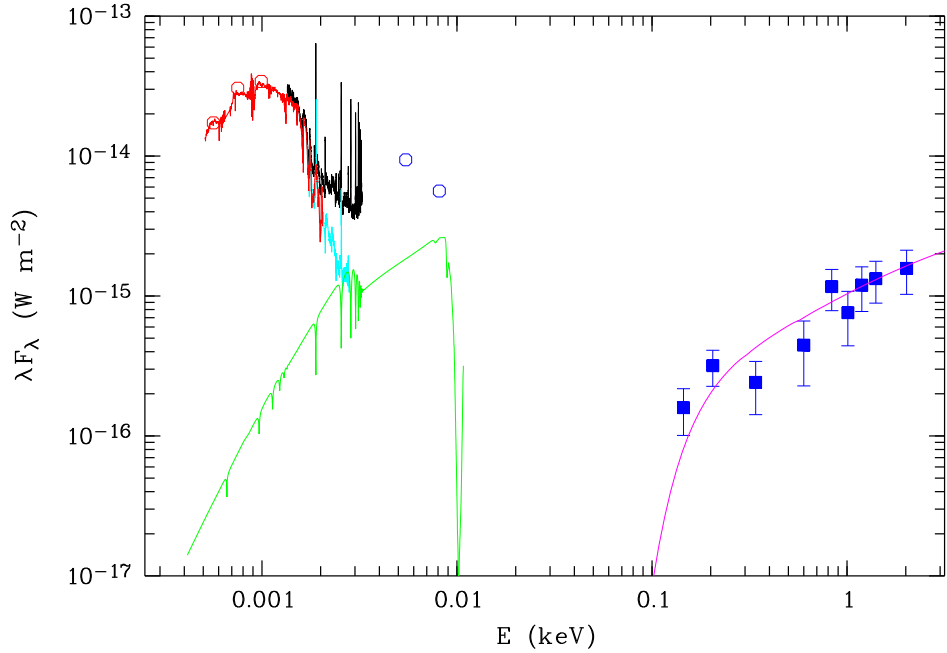}}
\caption{
Infrared to X-ray spectral energy distribution of V405 Peg. Shown are
in red 2MASS JHK photometric data points and a scaled version of the M3.5
dwarf LHS4 \citep{leggettetal00}. The optical high-state spectrum from SDSS
is shown in black, the optical low-state spectrum obtained by us in cyan.
A 13000\,K white dwarf model is shown in green. GALEX ultraviolet photometry
in FUV and NUV bands is shown with blue circles. The ROSAT RASS X-ray spectrum
(blue squares) was fitted with a thermal plasma component ($>3$\,keV),
slightly absorbed by cold interstellar
matter (magenta solid lines). 
{\it (see the online edition of the journal for a color version).}
}
\label{f:sed1955}
\end{figure}

\end{document}